\documentclass[letterpaper,twocolumn,prl,aps,superscriptaddress,floatfix]{revtex4}%,
\usepackage{lipsum} % 调用跨栏长公式宏包
\usepackage[latin9]{inputenc}
\usepackage{amsmath}
\usepackage{amssymb}
\usepackage{graphicx}
\usepackage{esint}
\usepackage{times}

\usepackage{color}

\usepackage[unicode=true,
bookmarks=true,bookmarksnumbered=false,bookmarksopen=false,
breaklinks=false,pdfborder={0 0 1},backref=false,colorlinks=true]
{hyperref}
\hypersetup{
    linkcolor=magenta, urlcolor=blue, citecolor=blue, pdfstartview={FitH}, hyperfootnotes=false, unicode=true}
    
\usepackage{afterpage}

\makeatletter

\pdfpageheight\paperheight
\pdfpagewidth\paperwidth

\@ifundefined{textcolor}{}
{%
    \definecolor{BLACK}{gray}{0}
    \definecolor{WHITE}{gray}{1}
    \definecolor{RED}{rgb}{1,0,0}
    \definecolor{GREEN}{rgb}{0,1,0}
    \definecolor{BLUE}{rgb}{0,0,1}
    \definecolor{CYAN}{cmyk}{1,0,0,0}
    \definecolor{MAGENTA}{cmyk}{0,1,0,0}
    \definecolor{YELLOW}{cmyk}{0,0,1,0}
}

\usepackage{xcolor}\usepackage{soul}

\definecolor{blue}{rgb}{0,0,1}
\definecolor{red}{rgb}{1,0,0}
\definecolor{green}{rgb}{0,1,0}

\makeatother

\begin{document}

\title{Decoherence-Suppressed Non-adiabatic Holonomic Quantum Computation}

\affiliation{Henan Key Laboratory of Diamond Optoelectronic Materials and Devices, Key Laboratory of Materials Physics of Ministry of Education,
	School of Physics and Microelectronics, Zhengzhou University, Zhengzhou 450052, China}
\affiliation{Department of Physics, Southern University of Science and Technology, Shenzhen, Guangdong 518055, China}

\author{B.-J. Liu \footnote{Present Address: Department of Physics, University of Massachusetts-Amherst, Amherst, MA, USA}}
\affiliation{Department of Physics, Southern University of Science and Technology, Shenzhen, Guangdong 518055, China}
\affiliation{Henan Key Laboratory of Diamond Optoelectronic Materials and Devices, Key Laboratory of Materials Physics of Ministry of Education,
	School of Physics and Microelectronics, Zhengzhou University, Zhengzhou 450052, China}

\author{L.-L. Yan}

\author{Y. Zhang}
\affiliation{Henan Key Laboratory of Diamond Optoelectronic Materials and Devices, Key Laboratory of Materials Physics of Ministry of Education,
	School of Physics and Microelectronics, Zhengzhou University, Zhengzhou 450052, China}
\author{M.-H. Yung}
\affiliation{Department of Physics, Southern University of Science and Technology, Shenzhen, Guangdong 518055, China}

\author{E.-J. Liang}
\author{S.-L. Su} \email{slsu@zzu.edu.cn}

\author{C. X. Shan}
\affiliation{Henan Key Laboratory of Diamond Optoelectronic Materials and Devices, Key Laboratory of Materials Physics of Ministry of Education,
	School of Physics and Microelectronics, Zhengzhou University, Zhengzhou 450052, China}
\date{\today}

\begin{abstract}
Nonadiabatic holonomic quantum computation~(NHQC) provides an essential way to construct robust and high-fidelity quantum gates due to its geometric features. However, NHQC is more sensitive to the decay and dephasing errors than conventional dynamical gate since it requires an ancillary intermediate state.
Here, we utilize the Hamiltonian reverse engineering technique to study the influence of the intermediate state-decoherence on the NHQC gate fidelity, and propose the novel schemes to construct the arbitrary single-qubit holonomic gate and nontrivial two-qubit holonomic gate with high fidelity and robustness to the decoherence.  Although the proposed method is generic and can be applied to many experimental platforms, such as superconducting qubits, trapped ions, quantum dots,  here we take nitrogen-vacancy (NV) center as an example to show that the gate fidelity can be significantly enhanced from 89\% to 99.6\% in contrast to the recent experimental NHQC schemes [Phys. Rev. Lett. 119, 140503 (2017); Nat. Photonics
11, 309 (2017); Opt. Lett. 43, 2380 (2018)], and the robustness against the decoherence can also be significantly improved. All in all, our scheme provides a promising way for fault-tolerant geometric quantum computation.

\end{abstract}

%\pacs{03.67.Lx, 42.50.Pq, 42.50.Dv}

\maketitle

{\it Introduction}.---Geometric quantum computation is based on that the time-dependent quantum state would obtain Abelian geometric phase~\cite{berry1984quantal,Aharonov1987Phase}, or non-Abelian holonomy~\cite{Wilczek1984,Anandan1988} via a cyclic quantum evolution.  
Since the holonomy (geometric phase) depends only on the global properties of the evolution paths and thus is robust to the local errors, it is important for constructing quantum gates for holonomic quantum computation (HQC)~\cite{Zanardi1999}.  The initial HQC~\cite{duan2001geometric,Pachos1999Nonabelian,sjoqvist2008trend,Wu2005Holonomic,single2,universal1} is based on adiabatic quantum evolution for realizing robust quantum operation against control errors~\cite{solinas2012on}. However, adiabatic HQC requires long operation time and thus it is sensitive to the environment-induced decoherence. To overcome such a problem, non-adiabatic holonomic quantum computation (NHQC)~\cite{Sjoqvist2012a,Xu2012Nonadiabatic} has been proposed to reduce the operation times of holonomic quantum gates~\cite{liu2019,Ramberg2019,Xue2015,Xue2017,three1,Composite2017holo,Hong2018,Zheng2016,Albert2016Holonomic,Zhao2020,Liu2020PRPP,GCY2020,wu2020error,YeHong2021,Liang2022}, and has also been demonstrated in recent years with different physical platforms including superconducting qubits~\cite{XXu2018prl,Abdumalikov2013,Egger2019,Yan2019,Han2020,Zhang2019,Danilin2018,li2021realization,xu2021demonstration}, nuclear magnetic resonance ~\cite{Feng2013,long2017,zhu2019}, trapped ions~\cite{ai2020experimental,ai2021experimental}, and nitrogen-vacancy~(NV)~centers in diamond~\cite{Zu2014,Arroyo2014,Nagata2018,sekiguchi2017optical,zhou2017holonomic,ishida2018universal,dong2021fast,dong2021experimental}, etc.  Although NHQC is robust to some certain errors, it requires an ancillary excited state, and thus is sensitive to the errors due to the decay and dephasing of this excited state  than the conventional dynamical gate. Up to now, in NV center system, the all-optical experiments show that the fidelity of NHQC gate are about 93\%~\cite{sekiguchi2017optical,zhou2017holonomic,ishida2018universal} , mainly limited by the decay and dephaisng errors.

In this paper, we propose an extensible approach of decoherence-suppressed NHQC (DS-NHQC), that  the decoherence of intermediate state can be greatly suppressed by using the Hamiltonian reverse engineering technology. Alternative: In this paper, we utilize the Hamiltonian reverse engineering technology to design the amplitude and phase of the laser pulse for the NHQC gate, and analyze the influence of the excited-state decay and dephasing on the fidelity of gate, and propose a series of novel schemes to realize the NHQC gates with high fidelity and robustness. In addition, universal holonomic gates including single- and two-qubit  gates, can be realized in a nitrogen-vacancy center system. Numerical simulations show that our DS-NHQC method can achieve a significant improvement of gate fidelities and robustness against decoherence over the NHQC gates in Refs.~\cite{sekiguchi2017optical,zhou2017holonomic,ishida2018universal} with the experimental parameters. Furthermore, our extensible approach of nondiabatic holonomic quantum computation is compatible  with decoherence-free subspace~\cite{Lidar1998} to further protect qubits from environmental noises.

\textit{General Model}.---Here, we consider a three-level system with the two-photon detuning, where the ground states $|0\rangle$ and $|1\rangle$ of the system are coupled to the excited state $|e\rangle$ by two pulses to realize the transitions of $|0\rangle \leftrightarrow|e\rangle$ and $|1\rangle\leftrightarrow|e\rangle$ with the detuning $\Delta(t)$, strength $\Omega_{0,1}(t)$ and phase $\phi_{0,1}(t)$. Using the rotating wave approximation and moving to the interaction frame, the Hamiltonian is given by ($\hbar\equiv1$ hereafter) $H(t)=\Delta(t)|e\rangle\langle e|+\sum_{k=0}^{1}\left(\frac{\Omega_{k}(t)}{2}e^{i\phi_{k}(t)}|k\rangle\langle e|+\mathrm{H.c.}\right)$. For our purpose, we assume that the Rabi strengths $\Omega_{0}(t)$ and  $\Omega_{1}(t)$ have the same time dependence, and they can be parameterized as  $\Omega_{0}(t)=-\Omega(t)\sin(\theta/2)$ and $\Omega_{1}(t)=-\Omega(t)\cos(\theta/2)$, where the ration of the two pulses  $\Omega_{0}(t)/\Omega_{1}(t)$ to be time-independent.
Consequently, the above Hamiltonian can then be expressed as
\begin{equation}\label{EQ1}
H(t)=\frac{1}{2}\left[\Omega(t) e^{i \phi_{1}(t)}|b\rangle\langle e|+{\rm H.c.}\right]+\Delta(t)|e\rangle\langle e|  \,
\end{equation}
where the time-independent bright state is defined by $|b\rangle\equiv-\sin(\theta/2)e^{i\phi}|0\rangle+\cos(\theta/2)|1\rangle$ with the phase $\phi=\phi_{0}-\phi_{1}$. Note that the "dark state" $|d\rangle\equiv\cos(\theta/2)|0\rangle+\sin(\theta/2)e^{-i\phi}|1\rangle$ is decoupled from the quantum dynamics between the subspace of $|b\rangle$ and $|e\rangle$.

With the help of quantum Householder reflections~\cite{householder1958unitary,Kyoseva2006,Genov2013prl}, the evolution operator $U(t)$ corresponding to Eq. (\ref{EQ1})  can be generally parameterized as
\begin{equation}\label{EQ2}
\begin{aligned}
U &=e^{i \alpha}\left(a^{*}|b\rangle\left\langle b\left|-b^{*}\right| b\right\rangle\langle e|+b| e\rangle\langle b|+a| e\rangle\langle e|\right)+|d\rangle\langle d| \\
&=\left(\begin{array}{ccc}
\mathrm{c}_{\frac{\theta}{2}}^{2}+a^{*} s_{\theta}^{2} e^{i \alpha} & b^{*} s_{\theta / 2} e^{i(\alpha+\varphi)} & \frac{e^{i \varphi}-a e^{i(\alpha+\varphi)}}{2} s_{\theta} \\
-b s_{\frac{\theta}{2}} e^{i(\alpha-\varphi)} & a e^{i \alpha} & b c_{\frac{\theta}{2}} e^{i \alpha} \\
\frac{e^{i \varphi}-a e^{i(\alpha-\varphi)}}{2} s_{\theta} & -b^{*} c_{\frac{\theta}{2}} e^{i \alpha} & a^{*} \mathrm{c}_{\frac{\theta}{2}}^{2} e^{i \alpha}+s_{\frac{\theta}{2}}^{2}
\end{array}\right) \ ,
\end{aligned}
\end{equation}
where $c_{x} \equiv \cos x$ and $s_{x} \equiv \sin x$.
Here, the time-dependent complex parameters $a$ and $b$ (with $|a|^{2}+|b|^{2}=1$) are the Cayley-Klein parameters of the SU(2) propagator in the effective two-state system of bright state $|b\rangle$ and excited state $|e\rangle$.  Without loss of generality, we can further parameterize the complex parameters as
$a=\cos \left(\frac{\eta}{2}\right)-i \sin \left(\frac{\eta}{2}\right) \cos \chi$ and $b=-i\sin \left(\frac{\eta}{2}\right) \sin \chi e^{-i\phi}$. Note that these parameters $\eta(t),\chi(t),\phi(t)$ and $\alpha(t)$ are generally time-dependent variables to be determined below. In this way, we rewrite the Eq. (\ref{EQ2}) in the original basis $\{|0\rangle,|e\rangle,|1\rangle\}$ as follows,
\begin{widetext} 
\begin{equation}\label{EQ3}
    U(t)=\left(\begin{array}{ccc}
c_{\theta / 2}^{2}+\left(c_{\eta / 2}+i s_{\eta / 2} \mathrm{c}_{\chi}\right) s_{\theta / 2}^{2} e^{i \alpha} & s_{\eta / 2} s_{\theta / 2} s_{\chi} e^{i(\pi / 2+\phi)} e^{i \alpha} e^{i \varphi} & \frac{1-\left(c_{\eta / 2}-i s_{\eta / 2} \mathrm{c}_{\chi}\right) e^{i \alpha}}{2} \mathrm{~s}_{\theta} e^{i \varphi} \\
-s_{\eta / 2} s_{\chi} s_{\theta / 2} e^{-i(\pi / 2+\phi)} e^{i \alpha} e^{-i \varphi} & \left(c_{\eta / 2}-i s_{\eta / 2} \mathrm{c}_{\chi}\right) e^{i \alpha} & s_{\eta / 2} s_{\chi} e^{-i(\pi / 2+\phi)} c_{\theta / 2} e^{i \alpha} \\
\frac{1-\left(c_{\eta / 2}-i s_{\eta / 2} \mathrm{c}_{\chi}\right) e^{i \alpha}}{2} \mathrm{~s}_{\theta} e^{-i \varphi} & -s_{\eta / 2} s_{\chi} c_{\theta / 2} e^{i(\pi / 2+\phi)} e^{i \alpha} & \left(c_{\eta / 2}+i s_{\eta / 2} \mathrm{c}_{\chi}\right) c_{\theta / 2}^{2} e^{i \alpha}+s_{\theta / 2}^{2}
\end{array}\right) \ .
\end{equation}
\end{widetext}
Using the time-dependent Schr\"{o}dinger equation, $i \dot{U}(t)=H(t) U(t)$, the Hamiltonian $H(t)$ can be expressed as $H(t)=i \dot{U}(t) U^{+}(t)$.  Consequently, the time dependence of the control pulses
can be determined by the following coupled differential equation, 
\begin{equation}\label{EQ4}
\begin{aligned}
&\Omega(t)=\sqrt{4 \dot{\chi}^{2} \sin ^{4} \eta+(\dot{\eta} \sin \chi+\dot{\chi} \cos \chi \sin \eta)^{2}} \\
&\Delta(t)=\dot{\eta} \cos \chi-\dot{\chi} \sin \chi \sin \eta \\
&\phi_{1}(t)=\phi+\pi / 2-\xi \\
&\dot{\alpha}(t)=(\dot{\chi} \sin \chi \sin \eta-\dot{\eta} \cos \chi) / 2
\end{aligned} \ ,
\end{equation}
 where $\xi(t)=\arctan \left[(\dot{\eta} \sin \chi+\dot{\chi} \cos \chi \sin \eta) /\left(2 \dot{\chi} \sin ^{2} \eta\right)\right]$. To eliminate the undesirable coupling, we shall keep $\phi$ time-independent for convenience. 
 
\textit{Decoherence-suppressed NHQC}.---To construct non-adiabatic holonomic gates in the computational basis $\{|0\rangle,|1\rangle\}$, the time-dependent parameters should satisfy the following nonadiabatic holonomic and cyclic conditions (see Supplemental Material~\cite{SM} for details), 
\begin{equation}\label{EQ5}
    \sin \left[\frac{\eta(\tau)}{2}\right] \sin [\chi(\tau)]=0 \ ,
\end{equation}
where the Eq. (\ref{EQ5}) ensures the evolution in the subspace $\{|0\rangle,|1\rangle\}$ is purely geometric, with vanishing dynamic contributions at each moment of time~\cite{Sjoqvist2012a,Xu2012Nonadiabatic}.
A possible choice satisfying the condition in Eq. (\ref{EQ5}) would be $\eta(\tau)=2 k \pi$. In this way, the evolution operator in  Eq. (\ref{EQ3}) acting on the computational space $\{|0\rangle,|1\rangle\}$ can be simplified as,
\begin{equation}\label{EQ6}
    \begin{aligned}
U_{s}(\theta,\gamma,\varphi) &=\left(\begin{array}{ll}
c_{\theta / 2}^{2}+s_{\theta / 2}^{2} e^{i \gamma} & \frac{1-e^{i \gamma}}{2} \mathrm{~s}_{\theta} e^{i \varphi} \\
\frac{1-e^{i \gamma}}{2} \mathrm{~s}_{\theta} e^{-i \varphi} & c_{\theta / 2}^{2} e^{i \gamma}+s_{\theta / 2}^{2}
\end{array}\right) \\
&=e^{i \frac{\gamma}{2}} e^{-i \frac{\gamma}{2} \mathbf{n} \cdot \mathbf{\sigma}}
\end{aligned} \ ,
\end{equation}
where $\gamma=k\pi+\alpha(\tau)$ is geometric phase, $\mathbf{n}=(\sin\theta\cos\phi,\sin\theta\sin\phi,\cos\theta)$, $\mathbf{\sigma}$ are the Pauli matrices.  Eq. (\ref{EQ6}) describes a rotational operation around the $\bm{n}$ axis by a $\gamma$ angle, up to a global phase factor $e^{-i\frac{\gamma}{2}}$. As both $\mathbf{n}$ and $\gamma$ can take any value,  Eq. (\ref{EQ6}) denotes a set of universal single-qubit gates in the qubit subspace.

In the construction of our geometric gate, the parametrized parameters $\eta(t),\chi(t),\phi$ and $\alpha(t)$ only need to satisfy the constraint in Eq. (\ref{EQ5}). Consequently, our approach allows one to construct the fault-tolerant NHQC (F-NHQC) via optimizing these parameters, where the population of the excited state $|e\rangle$ can significantly be suppressed.

\begin{figure}[htbp]
\centering
\includegraphics[width=8.5cm]{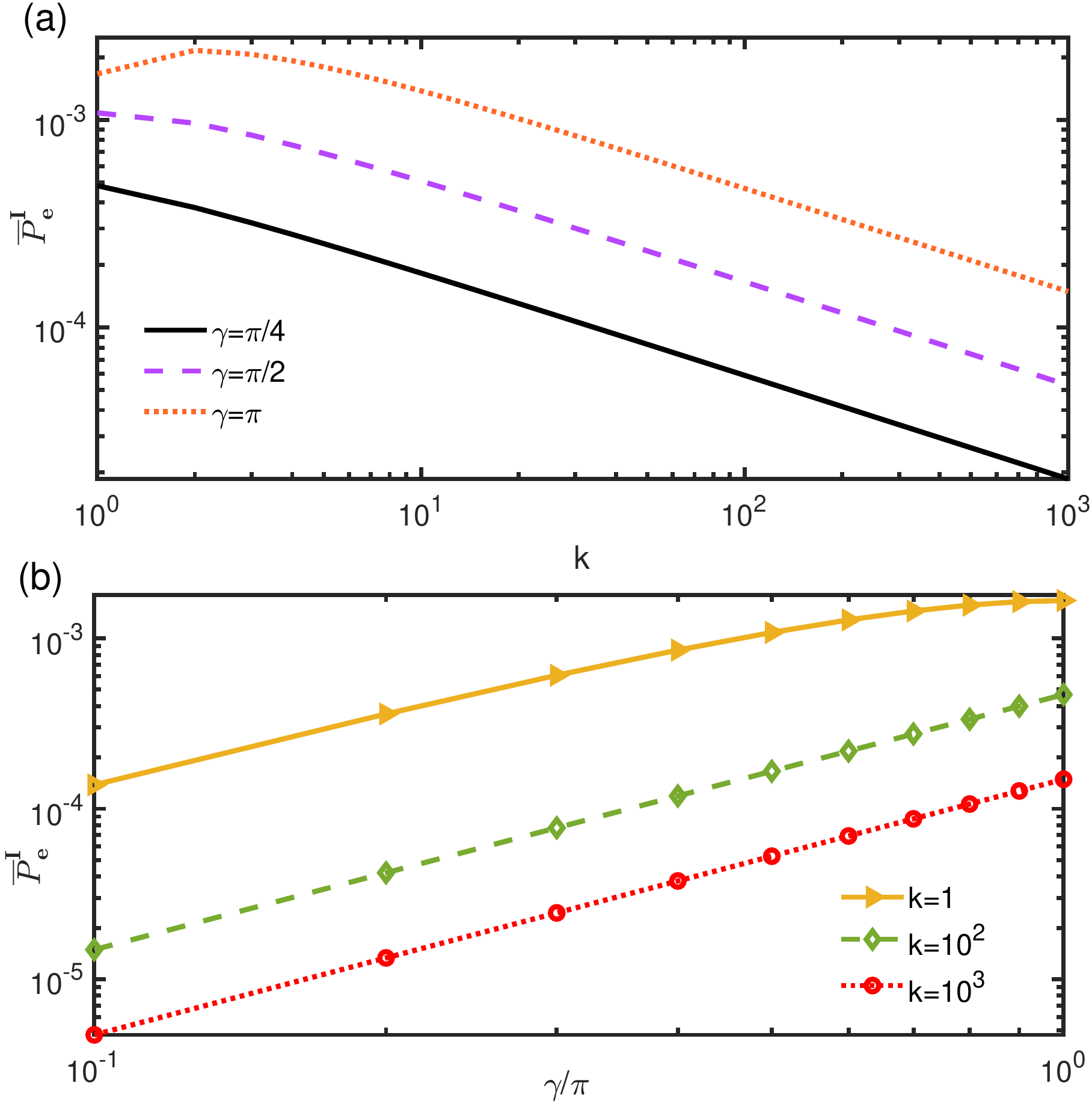}
\caption{ The average integrated population of excited state, $\bar{P}_{e}^{I}$, is taken as functions of (a) the integer $k$ and (b) the geometric phase $\gamma$.}\label{fig1}
\end{figure}

\textit{Suppressed Excited State Population}.---Here, we set a general initial state $\left|\psi_{i}\right\rangle=\sin \frac{\omega}{2}|0\rangle+\cos \frac{\omega}{2} e^{i \kappa}|1\rangle$, and then the integrated population of excited state defined by $P_{e}^{I} \equiv\int_{0}^{\tau}\left|\left\langle e|U(t)| \psi_{i}\right\rangle\right|^{2} d t$, is given by,
\begin{equation}\label{EQ7}
P_{e}^{I} =\left(\cos ^{2} \frac{\theta+\omega}{2}-\sin \theta \sin \omega \sin ^{2} \frac{\varphi+\kappa}{2}\right) f(\tau) \ ,
\end{equation}
where $f(\tau)\equiv\int_{0}^{\tau} \sin ^{2} \frac{\eta\left(t\right)}{2} \sin ^{2} \chi\left(t\right) d t$.  Note that the integrated population of excited state in Eq. (\ref{EQ7}), depends on the initial qubit state. By averaging over all the input states, the average integrated population of excited state $\bar{P}_{e}^{I}$ is calculated to be,
\begin{equation}\label{EQ8}
\bar{P}_{e}^{I} \equiv \frac{1}{4 \pi} \int_{0}^{\pi} \int_{0}^{2 \pi} P_{e}^{I} d \omega d \kappa=\frac{1}{8} f(\tau) \ .
\end{equation}
In order to suppress the excited state excitation, we only need to reduce the parameters $f(\tau)$. For the purpose, one simple optimization is to set the parameter $\chi$ time-independent, and minimize it. In this way, the time dependence of the control pulses in Eq. (\ref{EQ4}) can be simplified as $\Omega(t)=\dot{\eta}(t) \sin \chi, \Delta(t)=\dot{\eta}(t) \cos \chi$ with $\chi=\arccos{[-2\alpha(\tau)/\eta(\tau)]}=\arccos{(1-\gamma/k\pi)}$, and the average integrated population of excited state is taken as 
\begin{equation}\label{EQ9}
\bar{P}_{I}^{e}=\left(\frac{2 \gamma}{k \pi}-\frac{\gamma^{2}}{k^{2} \pi^{2}}\right) \int_{0}^{\tau} \sin ^{2} \eta(t) d t \ .
\end{equation}

\begin{figure}[t]
\centering
\includegraphics[width=8.5cm]{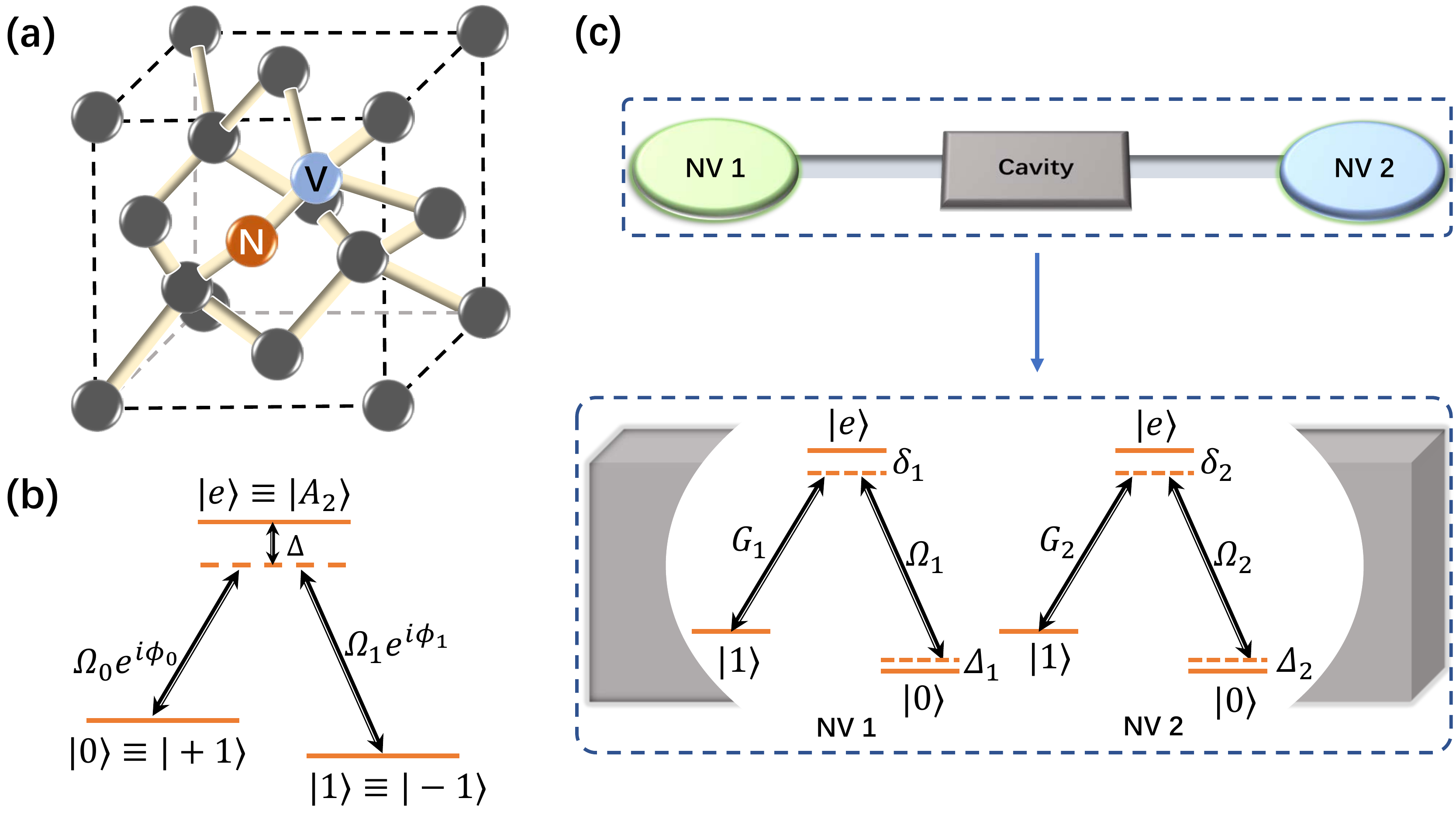}
\caption{Physical realization of DS-NHQC with NV centers. (a) Illustration of a NV centre in a diamond. (b) The level structure and coupling configuration for single-qubit operations in a optical NV $\Lambda$ system, where driven pulses with amplitudes $\Omega_{0,1}$ and phases $\phi_{0,1}$ couple the ground states $|0\rangle\equiv|+1\rangle$ and $|1\rangle\equiv|-1\rangle$ to  the  orbital  excited  state $|e\rangle\equiv|A_{2}\rangle$ with detuning $\Delta$. (c)  Illustration of a nontrivial two-qubit holonomic gate with two NV centers coupled to a cavity, and the crosponding coupling configuration illustrated in inset figure. }\label{setup}
\end{figure}

\begin{figure*}[htb]
\centering
\includegraphics[width=16cm]{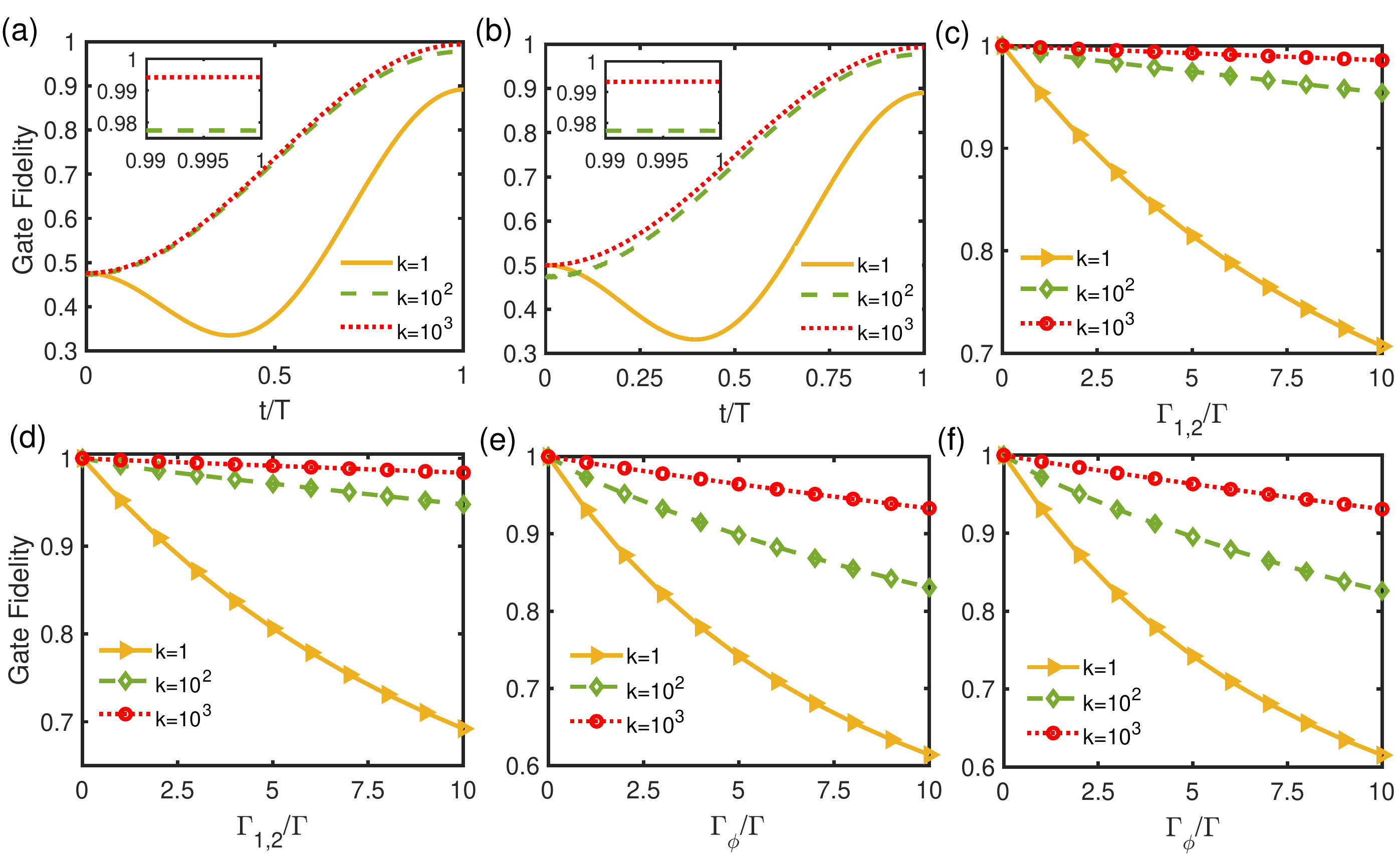}
\caption{ Illustration of the gate performance of DS-NHQC . Dynamics of gate fidelities of (a) NOT and (b)
Hardmard gates with different $k$. The NOT and Hardmard gate fidelities under the spontaneous decay error (c) and (d), orbital dephasing error (e) and (f), respectively.}\label{fig2}
\end{figure*}

Here, to illustrate the suppression of the intermediate state population, we plot the average integrated population of excited state $\bar{P}_{e}^{I}$ as functions of the integer $k$ with the parameters $\eta(t)=\Omega_{0} t$, and $\tau=2 k \pi / \Omega_{0}$, as shown in Fig. ~\ref{fig1}(a). The result clearly shows that one can choose the proper integer $k$ to reduce $\bar{P}_{e}^{I}$ to avoid unnecessary excited state population.  Moreover, the result shows that the average integrated population of excited state decreases with the decrease of the geometric phase $\gamma$ as shown in Fig. ~\ref{fig1}(b).

\textit{Example with NV centers.}---Here, we illustrate the implementation of our idea in realistic systems, namely, NV centers shown in Fig.~\ref{setup}(a).
Here we consider a three-level system in Refs.~\cite{yale2016optical,sekiguchi2017optical,zhou2017holonomic,ishida2018universal}, which spin coherence is controlled with light via the spin orbit interaction. The ground states $|\pm1\rangle$ of the NV are coupled to the orbital excited state $|A_{2}\rangle$ by two optical pulses to realize the transitions of $|+1\rangle \leftrightarrow|A_{2}\rangle$ and $|-1\rangle\leftrightarrow|A_{2}\rangle$ with one-photon detuning $\Delta(t)$, strength $\Omega_{0,1}(t)$ and phase $\phi_{0,1}$, shown in Fig. \ref{setup}(b). For our purpose, we choose $|0\rangle\equiv|+1\rangle$,  $|1\rangle\equiv|-1\rangle$ and  $|A_{2}\rangle\equiv|e\rangle$. Using the rotating wave approximation and moving to the interaction frame, the Hamiltonian is approximately given by Eq. (\ref{EQ1}). Therefore, we can realize arbitrary single-qubit gate of DS-NHQC with a NV device.

In an actually experimental system, the effects caused by the decay and dephasing process are unavoidable. To evaluate the performance of a holonomic gate in Eq.~(\ref{EQ6}) with the consideration of these effects, we use a quantum master equation in the Lindblad form~\cite{lindblad1976generators} as,
\begin{equation}\label{EQ10}
\dot{\rho}=-i[H, \rho]+\sum_{l=1}^{l=3}\left(L_{l} \rho L_{l}^{\dagger}-\frac{1}{2} L_{l}^{\dagger} L_{l} \rho-\frac{1}{2} \rho L_{l}^{\dagger} L_{l}\right) \ ,
\end{equation}
where $\rho$ is a density matrix of the considered NV center,  and $L_{l}$ is the Lindblad operators that denotes different dissipative effects in the realistic experiment. Specifically, $L_{1}\equiv|0\rangle\langle e|$ and $L_{2}\equiv|1\rangle\langle e|$ are spontaneous decay from $|e\rangle$ to $|0\rangle$ and $|1\rangle$ with the rate $\Gamma_{1}$ and $\Gamma_{2}$. And $L_{3}\equiv|e\rangle\langle e|$ is orbital dephasing of the level $|e\rangle$ with the rate $2\Gamma_{\phi}$. In our simulation, we use the following parameters in current  experiments~\cite{sekiguchi2017optical,zhou2017holonomic,ishida2018universal}, $\Omega_{0}=2\pi\times 300$ MHz, $\Gamma_{1}=2\Gamma_{2}=\Gamma=2\pi\times 8$ MHz, and  $\Gamma_{\phi}=2\Gamma$. Here, we have investigated the gate fidelity $F_{N}$ ($F_{H}$) of the NOT (Hardmard) gate with different $k$ for initial states of the form $|\psi(0)\rangle=\cos\zeta|0\rangle+\sin\zeta|1\rangle$, where a total of 1001 different values of $\zeta$ were uniformly chosen in the range of $[0, 2\pi]$, as shown in Fig. \ref{fig2}(a) and \ref{fig2}(b). The obtained gate fidelities $F_{N}$ ($F_{H}$) of the NOT (Hardmard) gate are given by 97.71\% (97.75\%) and 99.42\% (99.34\%) with $k=100$ and $k=1000$, while the gate fidelities $F_{N}$ ($F_{H}$) of NHQC corresponding to $k=1$, are as low as $89.17\%$ and $89.01\%$.

We now proceed to show the robustness improvement of our model against the decoherence. Specifically,  to investigate the robustness against the spontaneous decay (orbital dephasing), we plot the NOT and H gate fidelities as a function of spontaneous decay and (orbital dephasing) rate $\Gamma_{1,2}$ ($\Gamma_{\phi}$) for DS-NHQC, and NHQC scheme applied in the recent experiments with NV centers~\cite{sekiguchi2017optical,zhou2017holonomic,ishida2018universal}. As shown in Figs.~\ref{fig2}(c-f), our schemes of DS-NHQC can both greatly suppress the spontaneous decay and orbital dephasing comparing with NHQC.

\emph{Two-qubit holonomic gate}.---We now shift our focus to the implementation of a two-qubit gate between two NV centers, based on the idea of the proposed decoherence-suppressed holonomic quantum processes. Specifically, we consider the two NV centers couple to a optical cavity with far detuning, as sketched in Fig. \ref{setup}(c). Here, the energy level of NV center is the same as the single-qubit case, and each NV center interacts with the driving field individually, and the strength of the $k$th NV interacting with the driving field is $\Omega_k$ ($k=1, 2$). The coupling between the two NVs is achieved by the cavity modes, which in each NV form a Raman-like process with the driving field. By setting the proper parameters, the cavity mode and driving field with frequency $\omega_c$ and $\omega_{d,k}$ are able to couple with the transition $\left\vert j\right\rangle \rightarrow \left\vert e\right\rangle$ with transition frequency $\omega_{ej,k}$ ($j=0,1$) of $k$th NV center, respectively. The coupling strength of the $k$th NV with the cavity mode and the driving field are $G_k$ and $(-1)^k\Omega_k$, respectively, and the coupling between the driving field and the two NVs has a $\pi$ phase difference. The detuning $\delta_k=\omega_{e0,k}-\omega_c =\omega_{e1,k}-\omega_{d,k}-\Delta_{k}$ are the same for each NV center. In the interaction picture, under the rotating wave approximation, the Hamiltonian is given by $H_{I}=\sum_{k=1}^{2}\left[\left(G_{k} a \sigma_{e 0, k}+\Omega_{k} \sigma_{e 1, k}\right) e^{i \delta_{k} t}+ {\rm H.c.}+\Delta_{k} \sigma_{11, k}\right]$ with $\sigma_{mn,k}=\sigma^{+}_{nm,k}\equiv|m\rangle_{k}\langle n|$, where $a^{\dag}(a)$ is the creation (annihilation) operator of the cavity mode. Under the large detuning condition, $|\delta_{k}|\gg |G_{k},\Omega_{k}|$, the effective Raman Hamiltonian can be written as $H_{e}=\sum _{k=1}^{2}( g_{k}a\sigma _{01}^{+}+{\rm H.c.})+\Delta\sigma_{11,k}$ with the detuning $\Delta_{k}=\Delta$, in which the effective cavity-assisted coupling strength $g_{k}= (-1)^{k+1}G_k\Omega_k/\delta_k$ can be conveniently tuned via modulating the amplitude and phase of the corresponding external driven laser field $\Omega_k$. In this way, the effective Hamiltonian can be obtained in the subspace $\{|100\rangle,|010\rangle,|001\rangle,|110\rangle,|101\rangle,|011\rangle\}$ as
\begin{equation}\label{EQ11}
\begin{aligned}
H_{e}&=\left(\tilde{g}\left|B_{1}\right\rangle\langle 010|+\text { {\rm H.c.} }\right)-\Delta|010\rangle\langle 010| \\
&+\left(\tilde{g}\left|B_{2}\right\rangle\langle 101|+\text { {\rm H.c.} }\right)+\Delta|101\rangle\langle 101| \ ,
\end{aligned}
\end{equation}
where the abbreviation $|mnl\rangle \equiv|m\rangle_{1} \otimes|n\rangle_{c} \otimes|l\rangle_{2}$ denotes kronecker product of the state of the first NV center, the cavity mode, as well as the second NV center, respectively. The effective coupling strength in Hamiltonian~(\ref{EQ11}) is $\tilde{g}=\sqrt{g_{1}^{2}+g_{2}^{2}}$. And the two time-independent bright states are defined by $\left|B_{1}\right\rangle=-s_{\Theta / 2}|100\rangle+c_{\Theta / 2}|001\rangle$ and $\left|B_{2}\right\rangle=-s_{\Theta / 2}|110\rangle+c_{\Theta / 2}|011\rangle$ with mixing angle $\Theta=\tan ^{-1}\left(-g_{1} / g_{2}\right)$.

Similar to the single-qubit case, the nontrivial two-qubit entangled gate in the computational basis $\{|000\rangle,|100\rangle,|001\rangle,|101\rangle\}$ is given by
\begin{equation}\label{EQ12}
    U_{E}=\left(\begin{array}{cccc}
1 & 0 & 0 & 0 \\
0 & c_{\Theta / 2}^{2}+s_{\Theta / 2}^{2} e^{i \gamma} & s_{\Theta}\left(1-e^{i \gamma}\right) / 2 & 0 \\
0 & s_{\Theta}\left(1-e^{i \gamma}\right) / 2 & c_{\Theta / 2}^{2} e^{i \gamma}+s_{\Theta / 2}^{2} & 0 \\
0 & 0 & 0 & e^{-i \gamma} 
\end{array}\right) \ .
\end{equation}
On the basis of Eq. (\ref{EQ12}), a SWAP-like gate can be obtained by choosing the parameters $\Theta=\pi/2$ and $\gamma=\pi$. By combining the two-qubit gate $U_{E}$ and the single-qubit gates $U_{s}$, one can realize a universal set of decoherence-suppressed quantum gates.

\emph{Conclusion}.---We have presented a novel framework of decoherence-suppressed nonadiabatic geometric quantum computation, DS-NHQC, where the decay and dephasing noises of intermediate state can be greatly suppressed. Moreover, we have presented an explicit way to implement single- and two-qubit holonomic gates of DS-NHQC using the nitrogen-vacancy centers. Numerical results show our approach can significantly improve the gate fidelity and robustness of holonomic gates comparing with the recent experimental works of NHQC. Given its generality and simplicity, the DS-NHQC scheme can also be implemented on other platforms such as superconducting qubits, trapped ions, quantum dots, Rydberg atoms, and nuclear magnetic resonance, etc. Our work thus paves a way to construct fault-tolerant geometric quantum computation.

\acknowledgments
We are grateful to Prof. K. M{\o}lmer for his constructive suggestions on the manuscript.
This work is supported by the  National  Natural  Science  Foundation  of  China  (NSFC)  under  Grant No.11804308, U21A20434 and 12074346; Natural Science Foundation of Henan Province under Grants No. 212300410085and and China  Postdoctoral  Science  Foundation (CPSF)  under  Grant  No.2018T110735.

\bibliographystyle{Zou}
\bibliography{refs}

\end{document}